      [1999/12/01 v1.4c Il Nuovo Cimento]
\documentclass{cimento}

\usepackage{graphicx}
\usepackage{amssymb}
\title{The Beaming Factor and Other Open Issues in GRB Jets}
\author{Tsvi Piran\from{ins:x}\from{ins:y}}
\instlist{\inst{ins:x} Racah Institute for Physics, The Hebrew
University, Jerusalem, 91904, Israel,
  \inst{ins:y} Theoretical Astrophysics, Caltech, Pasadena, CA 91125, USA}
\PACSes{\PACSit{98.70.Rz ,95.85.Pw,95.30.Lz }{Gamma ray bursts,
gamma-rays,Hydrodynamics } }
\begin{document}

\maketitle

\begin{abstract}
I review several central open questions concerning GRB Jets.
\begin{itemize}
\item~ I discuss  a new estimate of the beaming correction for the
rate of  GRBs $\sim 75\pm 20$. \item ~ I discuss the universal
structured jet (USJ) model and conclude that while jets might be
structured they are less likely to be universal. \item~ I discuss
recent observations of a sideways expansion of a GRB afterglow and
compare these with current hydrodynamics simulations of jet
evolution. \item ~ I discuss the implications of resent outliers
to the energy-angle relation.
\end{itemize}
\end{abstract}

\section{Introduction}
The realization that the relativistic outflow in GRBs is in the
form of jets has direct implication to their energy budget and
their rates. This has, in turn, further implications on the nature
of the inner engines. For example prior to the understanding that
GRBs are beamed, events such as GRB 990123 with an isotropic
equivalent energy of more than $10^{54}$ erg were hard to explain.
With beaming the energy output of this event is a``mere" $10^{51}$
erg and it is compatible with a simple stellar mass progenitor.

Evidence of jetted GRBs arises  from observations
\cite{Harrisonetal99,Staneketal99} of the predicted achromatic
breaks in the afterglow light curves \cite{Rhoads99, SPH99}.
Further support is given by the comparison of long term radio
calorimetry with the energy of the prompt emission
\cite{Waxmanetal98}. The time of the jet break provides an
estimate of the jet angle \cite{SPH99}:
\begin{equation}
\theta =0.16 {\rm rad }(n/E_{k,iso,52})^{1/8} t_{b,days}^{3/8} =
0.07 {\rm rad } (n/E_{k,\theta,52})^{1/6} t_{b,days}^{1/2},
\label{eq:time}
\end{equation}
where $t_{b,days}$ is the break time in days and $E_{k,iso,52}$ is
the ``isotropic equivalent" kinetic energy in units of
$10^{52}$ergs, while $E_{k,\theta,52}$ is the real kinetic energy
in the jet i.e: $E_{k,\theta,52}=(\theta^2/2) E_{k,iso,52}$.

Frail et al. \cite{Frail01} and Panaitescu and Kumar
\cite{PanaitescuK01} have estimated the opening angles $\theta$
for several GRBs  with known redshifts. They find that the total
gamma-ray energy release, when corrected using a beaming factor,
$f_b$
\begin{equation}
E_\gamma = f_b E_{\gamma,iso} \equiv {\theta^2 \over 2 }
E_{\gamma,iso} ,
\end{equation} is
clustered. This energy-angle relation is commonly called the Frail
relation. A precursor of this discovery was found already in 1999
by Sari, Piran \& Halprn \cite{SPH99} who found that the two
brightest bursts known at that time were  beamed. Bloom, Frail \&
Kulkarni \cite{BloomFrailKulkarni03} confirmed this clustering
around $\sim 1.3\times 10^{51}$ erg on a larger sample. This
result is remarkable as it involves two seemingly unrelated
quantities, $\theta$ that is determined from the break in the
afterglow light curve and $E_{\gamma,iso}$ which is a property of
the prompt emission. The fact that the product of these two
unrelated quantities is a constant is, in my mind, an indication
that our overall model is correct.

There are two leading models for the jet structure and for the
interpretation of the jet break and the beaming angle. According
to the original uniform jet model (UJ) the energy per solid angle
is roughly constant within some finite opening angle, $\theta$,
and it sharply drops outside  $\theta$. Within the UJ model, the
observed break corresponds to the jet opening angle, $\theta$ (see
Eq. \ref{eq:time}). The Frail relation implies, here, that the
total energy released in GRBs is constant and that the differences
in the isotropic equivalent energies arise from variations in the
opening angles.

According to the alternative universal structured jet (USJ) model
\cite{Lipunov_Postnov_Pro01,Rossi02,Zhang02} all GRB jets are
intrinsically identical. The energy per solid angle varies as a
function of the angle from the jet axis. The jet break corresponds
to the viewing angle of the observer and the Frail relation
imposes a specific  distribution of energy per unit angle in the
jet:
\begin{equation}
\label{eq:usj} {\cal E}(\theta)=\left\{%
\begin{array}{ll}
    E_0/ (\pi \theta^2) , & \hbox{for $\theta>\theta_c$;} \\
    E_0/(\pi \theta_c^2), & \hbox{for $\theta<\theta_c$,} \\
\end{array}%
\right.
\end{equation}
where $E_0$ is a constant and $\theta_{c}$ is the core angle of
the jet \cite{Rossi02}. While the UJ model contains a free
function, the luminosity function or the corresponding opening
angle distribution, the USJ model is complectly determined by the
Frail relation. Its apparent luminosity function has the form
$\Phi(L)\propto L^{-2}$ \cite{Pernaetal03}. We can use this lack
of freedom within the USJ model to test it.

The question which model is the correct one is still an open one.
There are many other open questions concerning the structure of
the jets and their evolution. In this talk I review some of these
open questions. I refer the reader to a recent review \cite{p04}
for a more extended overview on GRBs in general and on GRB jets in
particular.

\section{The Beaming and the Rate of GRBs}

The overall GRB rate depends clearly on the amount that GRBs are
beamed. Within the UJ model this has  been measured traditionally
in terms of the  beaming correction factor, $f_b^{-1}$,  which is
defined as the ratio of total number of  bursts to the observed
ones. To estimate the overall GRB rate we need the average beaming
correction $\langle f_b^{-1} \rangle $ such that $n_{true}=
\langle f_b^{-1} \rangle n_{obs}$.  The average is performed over
the observed distribution. Taking into account the fact that for
every observed burst there are $f_b^{-1} $ that are not observed,
Frail et al., \cite{Frail01} estimated the average beaming
correction\footnote{Frail et al., \cite{Frail01} estimate the
observed beaming factor distribution as $p_{obs}(f_b) =
(f_b/f_0)^{\alpha+1}$ and $(f_b/f_0)^{\beta+1}$ for $f_b\lessgtr
f_0$ respectively. They find that $\alpha$ is poorly constraint by
the data while $\beta =-2.77_{-0.3}^{+0.24}$ and $log(f_0)=
-2.91_{-0.06}^{+0.07}$ and obtain $\langle f_b^{-1} \rangle_{F01}$
by integrating over this distribution.}
\begin{equation}
\langle f_b^{-1} \rangle_{F01} = {1\over N} \sum_i {2 \over
\theta_i^{2}} \simeq 520\pm 85 ,
\end{equation}
where the sum is over the observed distribution.

However, this calculation overestimates  the actual beaming
correction. In the {\it intrinsic} luminosity distribution there
are many low luminosity bursts that have large opening angles.
These bursts dominate the rate estimate. However, they are rather
weak and can be observed only to small distances. Hence they are
under-represented in the observed distribution \cite{gpw05}. This
effect can be taken into account in the following way. For a given
burst with a luminosity $L$ we define the volume from which such a
burst can be detected:
\begin{equation}
V_L \equiv \int_0^{z(L)} dz\,(dV/dz)R_{GRB}(z)/[R_{GRB}(0)(1+z)],
\end{equation}
where $R_{GRB}(z)$ is the comoving rate of GRBs and $z_m(L)$ is
the maximal redshift from which a burst with a luminosity L can be
detected. Similarly $V_\infty=V_{L=\infty}$  is the whole
effective volume of the observable universe. The  {\it intrinsic}
beaming correction can be written as:
\begin{equation}
\langle f_b^{-1} \rangle\equiv  {\sum_i (2 \theta_i^{-2})
(V_\infty/V_{L_i}) \over \sum_i (V_\infty/V_{L_i})} .
\end{equation}
This estimate is, of course, somewhat model dependent  as it
requires an assumption on $R_{GRB}$.  Guetta, Piran \& Waxman
\cite{gpw05} find that for several models in which GRBs follow the
SFR $\langle f_b^{-1} \rangle_{int} =75 \pm 25$, about a factor of
8 smaller than the previous estimate of $520\pm 85$ that did not
take this effect into account.

Following \cite{gpw05} we can also estimate the beaming correction
for the rate of GRBs within the  USJ model in the following way.
The total flux of GRBs per year (or per any other unit of time) is
an observed quantity obtained by summing over the observed
distribution. A comparison of this total flux with the total
energy emitted by a single burst can tell us directly the total
number of bursts. Integrating over the energy distribution (Eq.
\ref{eq:usj}) we obtain the total energy that a burst with a USJ
emits:
\begin{equation}
E_{\rm USJ} = 2  [\int_0^{\theta_c}(E_0/\theta_c^2)^2 \theta
d\theta + \int_{\theta_{c}}^{\theta_{\rm max}} E_0 \theta^{-1}
d\theta] = E_0  [1+2 \log ( \theta_{\rm max}/\theta_c) ],
\end{equation}
where $\theta_{max}$ is the maximal angle to which the jet
extends. This immediately implies that the ratio of UJs (emitting
each $E_0$)  to USJs (emitting each $E_0  [1+2 \log ( \theta_{\rm
max}/\theta_c) ]$) required to explain the observed flux is
\begin{equation}
N_{\rm UJ} /N_{\rm USJ} =  [1+2 \log ( \theta_{\rm max}/\theta_c)
].
\end{equation}
The upper and lower limits of this integral are uncertain but the
logarithmic dependance implies that the factor  cannot be smaller
than 2 or much larger than 5. This implies that  the number  of
USJs required to produce the observed flux is about factor of 4
below the corresponding number of UJs. Hence, the average beaming
correction for USJs is $\sim 20\pm 10$.

\section{Universal Structured Jets}

One of the intriguing open questions concerning GRB jets is their
angular distribution. The two leading models are the UJ and USJ
discussed earlier. The differences between USJ and UJ have crucial
implications to the question of the nature of GRBs' inner engines
and their progenitors. First, the universality of the USJ requires
that more or less the same process operates with the same
parameters within different GRBs. Second, a USJ carries roughly
five times more energy than a UJ. This implies, for a USJ,  an
energy budget of $\sim 10^{52}$ erg. The rate of USJs is,
correspondingly, smaller by a factor of five. It is therefore,
important to ask whether there are observations that can
distinguish between the two models.

\begin{figure}[b]
{\par\centering \includegraphics[width=10cm,height=10cm]{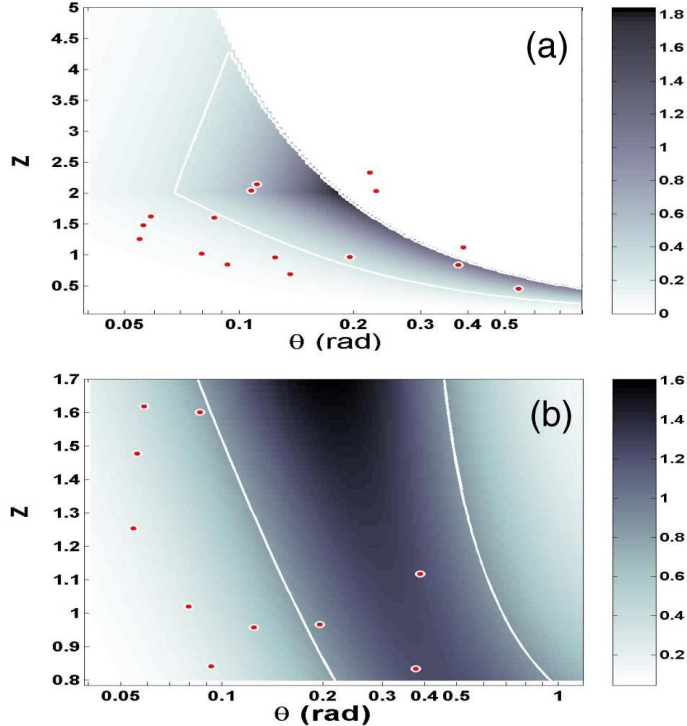}
\caption{{\bf (a)}: The 2D distribution density,
$dn(z,\ln\theta)/dz d\ln\theta$, of the GRB
  rate as a function of $z$ and $\ln\theta$ in the USJ model.
  The white contour lines confine the minimal
  area that contains $1\;\sigma$ of the total probability. The circles
  denote  $16$ bursts with  known $z$ and $\theta$ \cite{BloomFrailKulkarni03}.
  {\bf (b)}:  A limited redshift range,
  $0.8<z<1.7$   (containing $10$ out of the $16$ data points) in which both
  redshift selections effects and the sensitivity
  to the unknown GRB rate are minimized.  From \cite{ngg04}.}}
  \label{fig:1}
\end{figure}

Perna, Sari \& Frail \cite{Pernaetal03} calculated the expected
distribution of the observed opening angles within the USJ model
assuming that GRBs follow the SFR. Following them we define the
number of bursts in the interval $(\theta, \theta+d\theta)$ and
$(z, z+dz)$ as:
\begin{equation}
{dn\over d\theta dz} d\theta dz = sin \theta { R_{GRB}(z) \over
(1+z)} { dV(z) \over dz} {\cal T}(\theta,z) \label{eq:dndz}
\end{equation}
where, $V(z)$ is the comoving volume element and ${\cal
T}(\theta,z)$ depends on the distribution of GRB duration (see
\cite{Pernaetal03,ngg04} for details).  Perna et al.,
\cite{Pernaetal03} integrated over the redshift distribution and
found, with reasonable assumptions on ${\cal T}$ a remarkable
agreement between the expected angular distribution
$dn(\theta)/d\theta \equiv \int(dn/d \theta dz) dz$ and the
observed angular distribution, lending a strong support to the USJ
model. However, Nakar, Granot \& Guetta \cite{ngg04} compared of
the two dimensional distribution $dn/d\theta dz$ with the observed
one. They found (see Fig. \ref{fig:1}) that the two distribution
disagree strongly.  The observed points are very far from the
location of the peak of the expected distributions. Some are even
in a non-allowed region. This implies that the agreement between
the observed angle distribution and the one predicted by the USJ
model was just a coincidence and should not be taken as supporting
this model.

A similar discrepancy arose when Guetta et al., \cite{gpw05}
 compared the observed count ($C_{max}/C_{min}$) distribution for
the BATSE long bursts sample with the one expected from the USJ
model. Guetta et al., \cite{gpw05} found that the USJ model
predicts a significant  excess of weak bursts as compared with the
observed distribution.

\begin{figure}[b]
{\par\centering \resizebox*{0.7\columnwidth}{!}{\includegraphics
{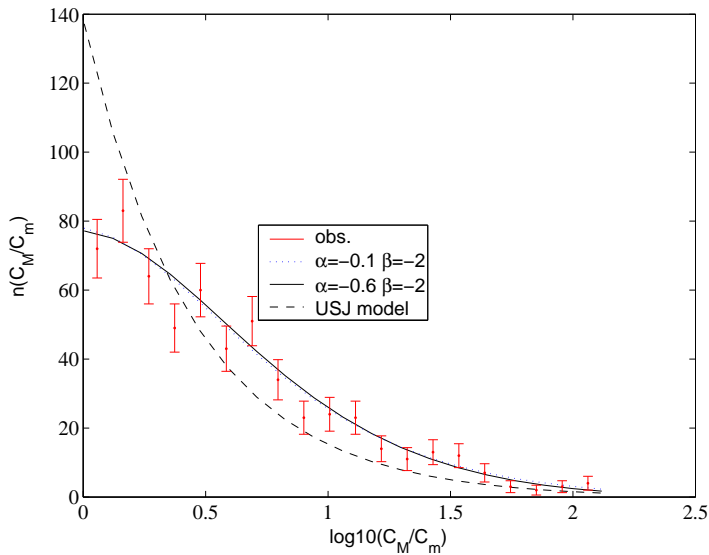}} \par}
 \caption{\label{fig:2} The predicted
differential $C_{\rm max}/C_{\rm min}$ distribution for the  USJ
model and the observed distribution taken from the BATSE catalog.
The inconsistency at the low peak flux range is apparent.
Predicted differential distributions for different
parameterizations of the luminosity function within the UJ model
are also shown. From \cite{gpw05}. }
\end{figure}

Selection effects and uncertainties mean that these discrepancies
are insufficient to rule out the USJ model. Still these
discrepancies certainly imply that the agreement found for the
angular distribution alone cannot be used, as hoped, to
demonstrate the validity of  this model. If it will turn out that
these conclusions hold with additional data, we will have to
reconsider the validity of the USJ. Both discrepancies could be
removed if we consider Structured Jets that are not universal.
That is if we allow an angular dependence of the flow parameters
(such as $\Gamma$ and/or $E$) but we do not require that all jets
are similar. Namely we could replace USJ with SJ. While such a
solution is viable it clearly takes away some of the simplicity
and the elegance of the USJ mode.

\section{Jet Evolution}

\begin{figure}[b]\begin{center}
\includegraphics[width=10cm,height=10cm]{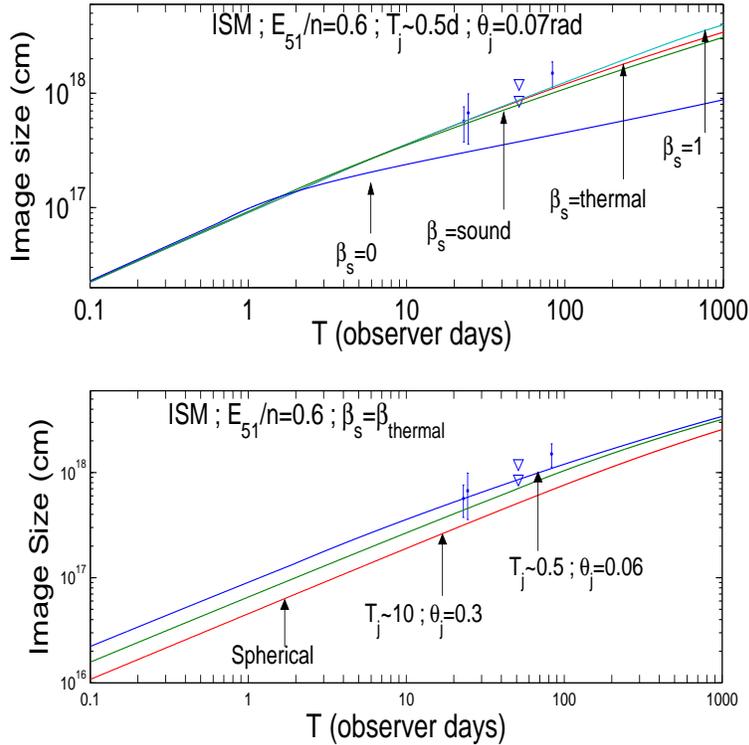}
\caption{{\it upper panel}: $R_\bot$ as a function of $T_d$ for
different sideways expansion models in ISM. The energy to external
density ratio $E/n=0.6\cdot 10^{51}$ erg \ cm$^3$ and
$\theta_0=0.06 rad$. {\it Lower panel}: $R_\bot$ as a function of
$T_d$ for different opening angles in ISM, with a constant
$E/n=0.6\cdot 10^{51}$ erg  cm$^3$. $T_j$ is in days,
$\beta_s=\beta_{thermal}$. From \cite{onp04} } \label{fig:3}
\end{center}
\end{figure}

An important question that determines the observed light curve of
the afterglow is the sideways expansion of the jet after $\Gamma
\sim \theta^{-1}$, where $\Gamma$ is the Lorentz factor. The
observations of the radio afterglow of GRB 030329 provided a
unique opportunity to test this issue. Taylor et al.,
\cite{tfbk04,tmp+04} have measured the size of the radio afterglow
of GRB 030329 between 20 and 300 days after the burst. Oren, Nakar
\& Piran \cite{onp04} (see also \cite{grl05}) compared the
observed sizes with several schematic models of spherical and
jetted propagation. The remarkable results is that the apparent
size of the afterglow is rather insensitive to the details of the
model. This robustness is a good indication for the validity of
the model, as it is difficult to force it to have different values
and the values obtained fit the observations. On the other hand it
is a drawback when one wishes to use the size to determine the
parameters of the outflow. It is insensitive to these parameters.

Oren et al., \cite{onp04} find that the image of a spherically
expanding fireball is largest with expansion as $t^{0.6}$, while
the size of a sideways expanding jet (at $v\approx c$) increases
as $t^{0.5}$. As can be seen in Fig. \ref{fig:3} there is
practically no difference between an expansion at the speed of
light or at the sound speed, $c\sqrt{3}$. On the other hand the
size of a non-expanding jet increases only as $t^{0.25}$. A
comparison of these models to the observations (see Fig.
\ref{fig:3}) shows that the non-expanding jet model is
inconsistent with the observations. While an addition of the
faster expansion during the Newtonian phase \cite{grl05} can
somewhat alleviate the problem, it is not clear that this can be
done with reasonable parameters \cite{onp04}.

Based on simple analytic model Sari Piran \& Halpern \cite{SPH99}
estimated that the relativistic jet will expand sideways almost at
the speed of light. On the other hand Panaitescu \& Meszaros
\cite{PanaitescuMeszaros99} estimated that the jet will expand
only at the sound speed, $c/\sqrt{3}$. Both are  consistent with
the observations. More recently Kumar \& Granot
\cite{KumarGranot03} integrated the hydrodynamic equations over
the radial direction and obtained one dimensional simplified
hydrodynamics equations. Solving these equations they find no or
little sideways expansion. Similar results were obtained in a  two
dimensional numerical integration of the full hydrodynamic
equations \cite{Granot01,Cannizzoetal04}. Remarkably this does not
influence much the observed light curve (see \cite{Granot01}).

We are left with a puzzle why do the numerical simulation indicate
little or no expansion while the observations suggest a rather
rapid sideways expansion. One may guess  that the current
computations are not refined enough to follow the jet evolution
(see \cite{pg01} for a detailed discussion of the problematics of
these computations). We probably have to wait for a higher
resolution codes to resolve this problem. Another possibility is
that the size of the radio afterglow does not trace well the size
of the expanding jet. Detailed emission computations have to be
carried out to determine this possibility.

\section{The Implication of Outliers to the Frail Relation and a Speculation}

We begun stressing the importance of the Frail (energy-angle)
relation. However, Berger et al., \cite{Bergeretal03_030329}
pointed out that in addition to the well known weak GRB 980425
there are three other outliers to this relation GRBs 980326,
980519 and 030329 . Intriguingly enough all outliers are weaker
relative to the common value of $\sim 10^{51}$ erg. I would like
to suggest a simple explanation to this phenomenon. It seems that
the common value indicates indeed the available energy reservoir.
Under optimal conditions a significant fraction of this energy is
released as $\gamma$-rays. An intriguing question, incidentally,
is how is this  conversion so efficient? These efficient cases
produce the brightest and easiest to detect bursts. The less
efficient case are less powerful (in $\gamma$-rays) and hence are
easily missed. GRB 980425 would not have been discovered if it was
not so close. Another outlier, GRB 030329, is also nearby at
z=0.168. The redshift of the other two outliers is unknown. Thus,
I conclude with a speculation that as time progresses and more
energies, redshifts and jet angles will become available it will
turn out that the Frail relationship is satisfied as an inequality
with $\sim 10^{51}$ erg being the upper limit to the $\gamma$-ray
energy.

\acknowledgments This work was supported by EU-RTN "GRBs Enigma
and a Tool" and by US-Israel BSF. I thank J. Granot, D. Guetta, E.
Nakar and E. Waxman for helpful discussions.


\begin{thebibliography}{10}

\bibitem{Harrisonetal99}
\BY{F.~A. {Harrison}, et al.  } \IN{Ap. J. Lett.}
{523}{1999}{L121--L124}.

\bibitem{Staneketal99}
\BY{K.~Z. {Stanek}, et al. } \IN{Ap. J. Lett.}
{522}{1999}{L39--L42}.


\bibitem{Rhoads99}
\BY{J.~E. {Rhoads}} \IN{ Ap. J.} {525}{1999}{737--749}.

\bibitem{SPH99}
\BY{R.~{Sari}, T.~{Piran}, and J.\~P. {Halpern}}
 \IN{ Ap. J. Lett.} {519}{1999}{L17--L20}.


\bibitem{Waxmanetal98}
\BY{E.~{Waxman}, S.~R. {Kulkarni}, and D.~A. {Frail}}
 \IN{ Ap. J.} {497}{1998}{288} .

\bibitem{Frail01}
\BY{ D.~A. {Frail}, et al. }
 \IN{ Ap. J. Lett.} {562}{2001}{L55--L58}.


\bibitem{PanaitescuK01}
\BY{A.~{Panaitescu} and P.~{Kumar}}
 \IN{ Ap. J. Lett.} {560}{2001}{L49--L53}.

\bibitem{BloomFrailKulkarni03}
\BY{J.~S. {Bloom}, D.~A. {Frail}, and S.~R. {Kulkarni}}
 \IN{ Ap. J.}  {594}{ 2003}{674}

\bibitem{Lipunov_Postnov_Pro01}
\BY{ V.~M. {Lipunov}, K.~A.
{Postnov}, and M.~E. {Prokhorov}.} \IN{ Astronomy
Reports}{45}{2001}{236-240}.

\bibitem{Rossi02}
\BY{E.~{Rossi}, D.~{Lazzati}, and M.~J. {Rees}} \IN{Mon. Not. RAS}
{332}{2002}{945--950}.

\bibitem{Zhang02}
\BY{B.~{Zhang} and P.~{M{\' e}sz{\' a}ros}} \IN{Ap. J.}
{571}{2002}{876--879}.

\bibitem{Pernaetal03}
\BY{R.~{Perna}, R.~{Sari}, and D.~{Frail}} \IN{Ap.
J.}{594}{2003}{379--384}.

\bibitem{p04}
\BY{T. ~Piran} \IN{ Rev. Mod. Phys.} {76}{2004}{1143-1210}.

\bibitem{gpw05}
\BY{D.~{Guetta}, T.~{Piran}, and E.~{Waxman}}\IN{Ap.
J.}{619}{2005}{412--419}.


\bibitem{ngg04}
\BY{E.~{Nakar}, J.~{Granot}, and D.~{Guetta}}
 \IN{ Ap. J. Lett.} {606}{2004}{L37--L40}.

\bibitem{PanaitescuMeszaros99}
\BY{A.~{Panaitescu} and P.~{M{\' e}sz{\' a}ros}}
 \IN{ Ap. J.} {526}{1999}{707--715}.

\bibitem{KumarGranot03}
\BY{P.~{Kumar} and J.~{Granot}} \IN{ Ap. J.} {591}{2003}{1075}

\bibitem{Granot01}
\BY{J.~{Granot}, et al.}
 in \TITLE{ Gamma-ray Bursts in the Afterglow Era}, (2001) pp.~ 312--+.

\bibitem{Cannizzoetal04}
\BY{J.~K. {Cannizzo}, N.~{Gehrels}, and E.~T. {Vishniac}}
 \IN{ Ap. J.} {601}{2004}{380--390}.

\bibitem{tfbk04}
\BY{G.~B. {Taylor}, et al. } \IN{ Ap. J. Lett.}
{609}{2004}{L1--L4}.

\bibitem{tmp+04}
\BY{G.~B. {Taylor}, et al. } \IN { ArXiv Astrophysics e-prints},
{12483}{2004}{}.

\bibitem{onp04}
\BY{Y.~{Oren}, E.~{Nakar}, and T.~{Piran}}
 \IN{ Mon. Not. RAS,} {353}{2004}{L35--L40}.

\bibitem{grl05}
\BY{J.~{Granot}, E.~{Ramirez-Ruiz}, and A.~{Loeb}} \IN { Ap. J.}
{618}{2005}{413--425}.

\bibitem{pg01}
\BY{T.~{Piran} and J.~{Granot}} in \TITLE{ Gamma-ray Bursts in the
Afterglow Era}, (2001) pp. 300+.


\bibitem{Bergeretal03_030329}
\BY{E.~{Berger}, et al. } \IN{ Nature} {426}{2003}{154--157}.

\end{thebibliography}

\end{document}